# Observation of superconductivity in the noncentrosymmetric nodal chain semimetal $Ba_5In_4Bi_5$

Yuzhe Ma[1,2], Yulong Wang[1,2], Yuxin Wang[1,2], Soham Manni[3,4], Qisheng Lin[3,5], Linlin Wang[3,4], Kun Jiang[1,2], Sergey L. Bud'ko[3,4], Paul C. Canfield[3,4], Gang Wang[1,2,3,6*]

*1 Beijing National Laboratory for Condensed Matter Physics, Institute of Physics, Chinese Academy of Sciences, Beijing 100190, China*

*2 University of Chinese Academy of Sciences, Beijing 100049, China*

*3 Ames National Laboratory, Iowa State University, Ames, Iowa 50011, USA*

*4 Department of Physics and Astronomy, Iowa State University, Ames, Iowa 50011, USA*

*5 Department of Chemistry, Iowa State University, Ames, Iowa 50011, USA*

*6 Songshan Lake Materials Laboratory, Dongguan, Guangdong 523808, China*

The combination with superconductivity and topological nontrivial band structure provides a promising route towards novel quantum states such as topological superconductivity. Here, we report the first observation of superconductivity (4.1 K) in $Ba_5In_4Bi_5$ single crystal, a noncentrosymmetric topological semimetal featuring nodal chain loops at the high-symmetry points R and X. The magnetization, resistivity, and specific heat capacity measurements reveal that $Ba_5In_4Bi_5$ is a moderately coupled type-II Bardeen-Cooper-Schrieffer superconductor. Bulk superconductivity is suggested from the magnetic susceptibility and specific heat measurements. The results show that $Ba_5In_4Bi_5$ provides a new platform for exploring the relationship of superconductivity and topological nontrivial band topology.

---

* gangwang@iphy.ac.cn

The fascinating development in topological quantum matter during the last two decades has spurred a rush in exploring the emerging topological superconductivity (TSC) in condensed matter physics, which provides a promising avenue for the actualization of Majorana fermions. Two ways towards realizing TSC have been adopted: finding superconductivity (SC) with intrinsic nontrivial topology or combining the conventional SC with other nontrivial topological band structures. The search for intrinsic TSC, such as the $p_x+ip_y$ spin-triplet SC (e.g., $Sr_2RuO_4$, $UTe_2$) [1-4], has been very challenging. Recently, the searching for intrinsic TSC has been extended to superconducting topological materials via chemical doping or intercalation, e.g., the doped topological insulators Cu/Sr-doped $Bi_2Se_3$ [5-10], In-doped SnTe [11], and doped Weyl semimetals [12], which require fine tuning in composition and inevitably consist of defects. Besides, the TSCs have been also explored on the interface of a heterostructure between an s-wave superconductor and a strong topological insulator (e.g., $Bi_2Se_3/NbSe_2$, $Bi_2Te_3/NbSe_2$) [13,14], or fully-gapped bulk SC connate with topological surface states (e.g., $FeTe_{1-x}Se_x$, 2M-$WS_2$, and $TaSe_3$) [15-19] based on the superconducting proximity effect. Based on the discussion above, type-II superconductors with nontrivial surface states are promising candidates for TSCs. Therefore, it is highly desirable to explore new materials having both nontrivial band topology and intrinsic SC.

As a special nodal line, nodal chain consisting of connected loops that touch each other at isolated points on a high-symmetry axis, aroused the interest of Bzdušek et al. in 2016, and they proposed that such nodal chains could emerge in materials with the space groups of 102, 104, 109, 118, and 122 [20]. To this point, $Ba_5In_4Bi_5$ has attracted our attention: $Ba_5In_4Bi_5$ has been reported to be a Pauli paramagnetic semimetal crystalizing in a noncentrosymmetric tetragonal space group *P*4*nc* (No. 104) [21]. Very recently, it has been predicted to possess Weyl nodal line close to the Fermi energy [22]. In this work, we have successfully grown the topological nodal chain semimetal $Ba_5In_4Bi_5$ single crystals by a flux method and observed the SC for the first time, which is a type-II moderately coupled superconductor with a superconducting transition temperature ($T_c$) about 4.1 K. The in-plane and out-plane upper critical field shows weak anisotropy due to the three-dimensional nature of crystal structure. The observed superconducting shielding fraction and the jump of specific heat capacity at 4.1 K strongly suggest bulk superconductivity in $Ba_5In_4Bi_5$. The presence of intrinsic SC in a possible topologically nontrivial state makes $Ba_5In_4Bi_5$ a new platform to study the interplay of topological nodal chain and SC. In addition, the lack of the inversion symmetry in the structure might result in the discovery of noncentrosymmetric superconductor with mixed spin-singlet and spin-triplet SCs [23].

$Ba_5In_4Bi_5$ crystallizes in a tetragonal structure with the noncentrosymmetric space group *P*4*nc* (No. 104). The structure features isolated heteroatomic, square pyramidal $[In_4Bi_5]^{10-}$ anionic clusters that are separated by $Ba^{2+}$ cations. Each of the four In1 base atoms in the square pyramidal $[In_4Bi_5]^{10-}$ clusters is exo-bound to Bi1 atom in a distance of 2.9106(14) Å, in contrast, the base-to-apex In1-Bi2 distance of the pyramid is about 0.41 Å longer, 3.3212(19) Å (cf. Table S3). Moreover, adjacent square pyramids are shifted by *c*/2 and rotated by 16.4 degrees with respect to each other along the *c* axis, which reduces the overall symmetry of the crystal structure from body-centered to the tetragonal primitive unit cell. In addition, Bi2 has a similar coordination environment with that of Ba2, as can be seen from the right panel of Fig. 1(a). From Fig. S1(b), the neighboring pyramids are stacked vertically along the *c* axis and linked via Bi1 atoms and In1 atoms from different pyramids. The Bi1-In1 bonding distance is 3.3830(13) Å, which is close to the bonding distance of 3.3212(19) Å between Bi2 atoms and In1 atoms within the pyramids, so the interaction between the nearest neighboring clusters can't be ignored. What's more, $Ba_5In_4Bi_5$ is an electron-deficient compound. In electron-deficient compounds, Madelung energy and packing efficiency become more dominant than covalent bonding [21]. In order to further study this one-electron-deficient compound, we calculated the electronic band structure of $Ba_5In_4Bi_5$ with spin-orbit coupling (SOC) using the first-principles calculations, with details described in Supplemental Material. Figure S2(b) presents an overview of the band structure and density of states (DOS) of $Ba_5In_4Bi_5$. Close to the Fermi level, the major contribution to the DOS originates from Bi. A zoom-in view of the bands along the Γ-X-R path with SOC, shown in Fig. 1(b), reveals a band crossing along the Γ-X direction and a twofold degeneracy along the X-R direction, forming two Weyl nodal loops centered at R and X points. The band crossing and the twofold degeneracy can be clearly seen on the right of Fig. 1(b). The nodal loops touch each other, as displayed in Fig. 1(c) and discussed in reference [22]. Notably the crossing along the Γ-X direction of the Weyl nodal loop is very close and only 5 meV below the Fermi energy.

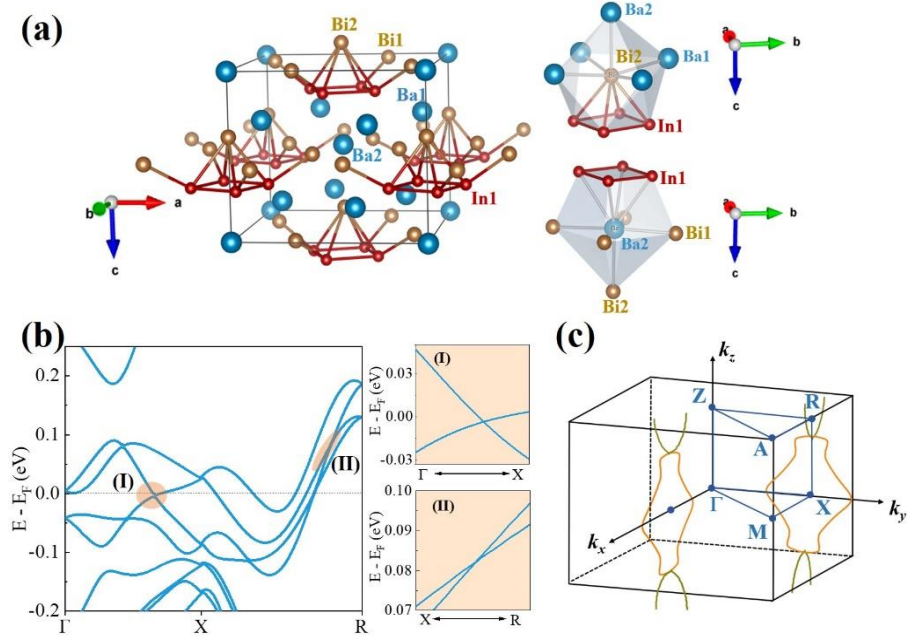

**FIG. 1.** (a) The schematic crystal structure of $Ba_5In_4Bi_5$ and the coordination environment of Bi2 and Ba2 in the unit cell. (b) The electronic band structure of $Ba_5In_4Bi_5$ along the Γ-X-R path with SOC. Right (I) and (II): Zoom-in band structures of crossing and twofold degeneracy. (c) The illustration of the Weyl nodal chain along the $k_z$ direction centered on the R and X points, respectively, which is formed by the energy bands in (b). Yellow: the nodal loops protected by the mirror plane as defined by $k_x = 0$ or $k_y = 0$. Green: the nodal loops protected by the mirror plane as defined by $k_x = π$ or $k_y = π$.

Inspired by the calculations, we grew $Ba_5In_4Bi_5$ single crystals by the flux method and studied their physical properties (See the supplementary material for details). Before we perform measurements, the crystals were stored in an argon-filled glove box to avoid sample degradation because, by visual inspection, they turn into black after exposure in air for 1 h. The crystal structure data of $Ba_5In_4Bi_5$ determined from single crystal X-ray diffraction are given in Table S1-S4. The lattice parameters of $Ba_5In_4Bi_5$ single crystal are $a$ = 10.6132(6) Å, $c$ = 9.0174(11) Å, and $Z$ = 2, which are in good agreement with those reported in the reference [21]. As shown in Fig. S1(c), the powder X-ray diffraction pattern of the crushed crystals matches well with the standard pattern in ICSD database, further confirming the crystal structure. The energy-dispersive X-ray spectroscopy (EDX) spectrum indicates that the atomic ratio of Ba, In, and Bi (35.57%:28.54%:35.89%) matches the stoichiometric ratio, and the elemental mapping shows that the chemical composition is quite uniform, as depicted in Fig. S1(d). More EDX data can be found in Table S5. Note that the inductively coupled plasma-atomic emission spectroscopy analyses exhibit the atomic ratio of Ba, In, and Bi in $Ba_5In_4Bi_5$ single crystal is close to 5:4:5, as displayed in Table S6, which is consistent with the EDX results. The Laue diffraction pattern of $Ba_5In_4Bi_5$ single crystal in Fig. S3 shows that the largest surface of a plate-like single crystal is (001) plane, which exhibits the fourfold symmetry.

Figure 2(a) shows the temperature-dependent magnetic susceptibility in the zero-field cooling (ZFC) and field cooling (FC) modes for a $Ba_5In_4Bi_5$ crystal with dimension of 1.4 × 0.5 × 0.25 mm$^3$ under a magnetic field of 10 Oe parallel to the (001) plane. The $T_c$, defined as the temperature at which the steepest slope of the superconducting signal in the ZFC data intersects the extrapolation of the normal-state magnetic susceptibility to lower temperatures, is 4.15 K. To get the more accurate value of the superconducting shielding fraction, the demagnetization factors ($N$) for $H$ // (001) and $H \perp$ (001) are estimated to be 0.15 and 0.62 respectively, by using $N^{-1} = 1 + \frac{3}{4} * \frac{c}{a}(1 + \frac{a}{b})$, where $a$, $b$, and $c$ are the length, width, and thickness of a rectangular cuboid superconductor [23]. Taking the error bars of applied magnetic field into consideration, the superconducting shielding fraction, normalized by a demagnetization factor, is estimated to be about 48%-90% at 3 K under the magnetic field of 10 ± 3 Oe, combined with the non-saturated ZFC data at 3 K, indicating the bulk nature of the SC in $Ba_5In_4Bi_5$. In addition, the typical hysteresis loop at 3 K, as shown in Fig. 2(b), suggests $Ba_5In_4Bi_5$ is a type-II superconductor. Figure 2(c) and (d) show the magnetic-field-dependent magnetization ($M(H)$) at temperatures below $T_c$ for $H$ // (001) and $H \perp$ (001), respectively. The lower critical field ($H_{c1}^*$) is determined by the point of $\Delta M = M - M_{fit}$ = 0.05 emu cm$^{-3}$ ($H$ // (001)) or 0.03 emu cm$^{-3}$ ($H \perp$ (001)) for each temperature, as shown in Fig. S4. Taking into account the demagnetization factor, the real lower critical field ($H_{c1}$) can be deduced from the $H_{c1}^*$ with the formula $H_{c1} = H_{c1}^* / (1 - N)$. The estimated $H_{c1}^{//(001)}(0)$ = 92 Oe and $H_{c1}^{\perp(001)}(0)$ = 82 Oe, as displayed in Fig. 2(e) and (f), are obtained by fitting the experimental data after making demagnetization correction with the equation $H_{c1}(T) = H_{c1}(0)[1 - (T/T_c)^2]$. One more thing to note is that the bulges under low field in the hysteresis loop at 3 K for $H$ // (001) may be caused by magnetic flux pinning, which often occur in superconductors containing impurities, defects, or dislocations [24]. When $Ba_5In_4Bi_5$ crystal was exposed in air for 2 days, two superconducting transitions were observed (Fig. S6).

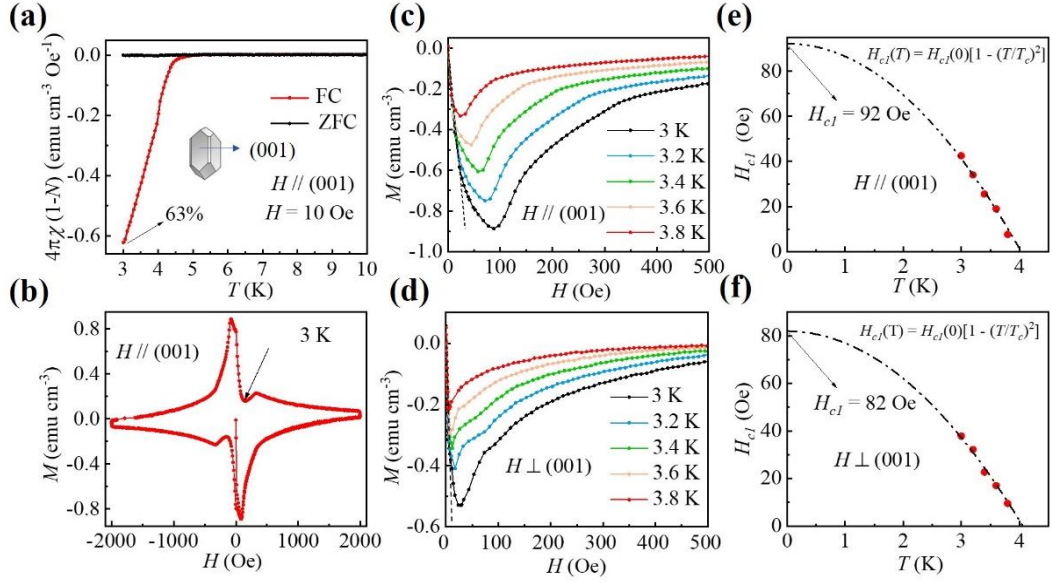

**FIG. 2.** (a) The ZFC and FC magnetic susceptibility for $Ba_5In_4Bi_5$ under a magnetic field of 10 Oe parallel to (001) plane. The absence of notable superconducting feature in the FC curve may be due to the magnetic flux pinning. (b) The hysteresis loop measured at 3 K. The black arrow denotes the bulge under low field. (c), (d) The $M(H)$ curves at different temperatures with $H // (001)$ plane and $H \perp (001)$ plane, respectively. (e), (f) The temperature dependence of the lower critical field after making demagnetization correction for $H // (001)$ plane and $H \perp (001)$ plane, respectively. The dotted lines show fittings using the equation $H_{c1}(T) = H_{c1}(0)[1 - (T/T_c)^2]$.

The temperature-dependent resistivity for $Ba_5In_4Bi_5$ single crystal measured in the temperature range of 2 K-300 K for $H \perp (001)$ with current applied in (001) plane (corresponding to the $ab$ plane of $Ba_5In_4Bi_5$ single crystal) is shown in Fig. 3(a). A metallic-like behavior ($d\rho/dT > 0$) is observed at relatively higher temperature ($T > 4.4$ K). The corresponding residual resistivity ratio (RRR) is around 6, indicating the good metallicity of the grown $Ba_5In_4Bi_5$. The inset of Fig. 3(a) displays the enlarged resistivity at low temperatures, in which a clear superconducting transition appears with

$T_c^{onset}$ of 4.4 K and $T_c^{zero}$ of 4.1 K. The low-temperature in-plane resistivity under different magnetic fields for $H \perp (001)$ plane and $H // (001)$ plane is presented in Fig. 3(b) and (d), respectively. As can be seen, $T_c$ decreases and the superconducting transition gradually broadens with increasing magnetic field. For $H \perp (001)$ plane, $T_c$ is not detectable above 2 K under magnetic field of 15 kOe, whereas zero resistivity is still reached above 2 K under magnetic field of 15 kOe for $H // (001)$ plane. $T_c$ is determined by 50% $\rho_n$ (the normal $\rho$ upon $T_c^{onset}$) here. The temperature-dependent upper critical field is plotted in Fig. 3(c) and (e). The curve can be well fitted by using the empirical equation $H_{c2}(T) = H_{c2}(0)(1 - t^{3/2})^{3/2}$ [25], where $t = T/T_c$. According to the equation, the estimated upper critical field is ~ 25.5 kOe and ~ 36.3 kOe for $H \perp (001)$ plane and $H // (001)$ plane, respectively. The upper critical fields for both configurations are much lower than the Pauli paramagnetic limit $H^{Pauli}(0) = 1.84T_c$ ~ 76 kOe [26]. Similar results have been observed in other single crystals (Fig. S7 and Fig. S8). We also used the Werthamer-Helfand-Hohenberg (WHH) expression $H_{c2}(0) = -AT_c dH_{c2}/dT|_{(T=T_c)}$ [27] and the Ginzburg-Landau (GL) equation $H_{c2}(T) = H_{c2}(0)(1 - t^2)/(1 + t^2)$ [28], where $t = T/T_c$, to fit the upper critical field, as shown in Fig. S9. Table S7 gathers $H_{c2}(0)$ values obtained from three different methods. In all cases, the anisotropy of $H_{c2}(0)$ is smaller than 2, which might be ascribed to the more three-dimensional crystal structure of Ba$_5$In$_4$Bi$_5$. The empirical equation gives an excellent fit ($R^2 = 0.99997$) compared with WHH and GL models, so we employ $H_{c2}(0) = 25.5$ kOe ($H \perp (001)$) and 36.3 kOe ($H // (001)$) for further calculations. All the calculated parameters are listed in Table S8.

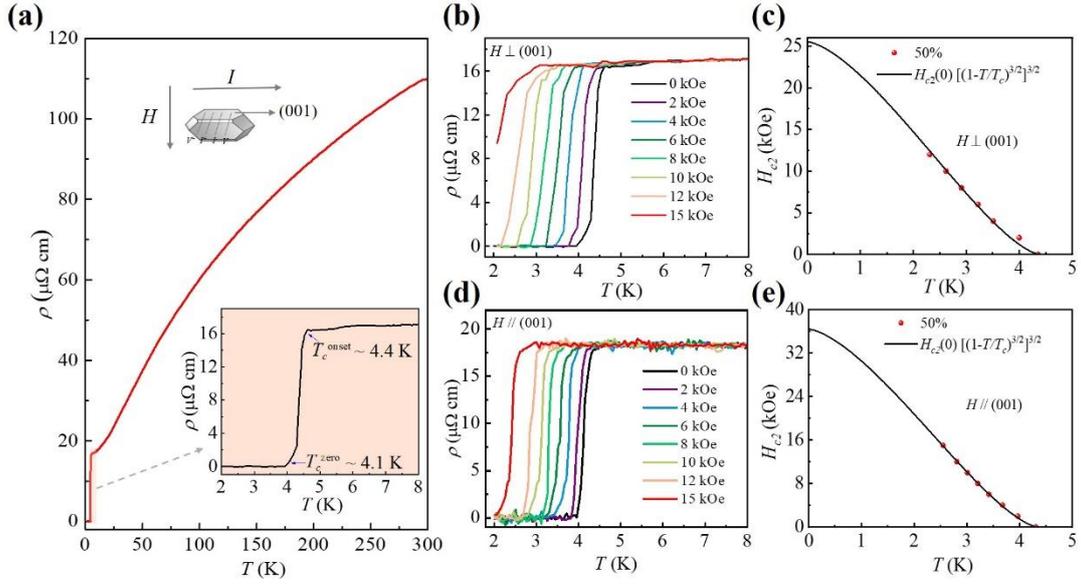

**FIG. 3.** (a) The temperature dependence of resistivity for $Ba_5In_4Bi_5$ single crystal with current applied in (001) plane for $H \perp$ (001) plane. The inset is the enlarged resistivity curve at low temperature range. (b), (d) The resistivity as a function of temperature under different magnetic fields for $H \perp$ (001) plane and $H //$ (001) plane, respectively. (c), (e) The temperature dependence of the upper critical field. The solid lines show fittings by the equation $H_{c2}(T) = H_{c2}(0)[1-(T/T_c)^{3/2}]^{3/2}$.

To further characterize the SC of $Ba_5In_4Bi_5$, the specific heat capacity $C$ was measured without magnetic field and with a magnetic field of 50 kOe for $H \perp$ (001) plane. Figure 4(a) displays the temperature-dependent $C$ ranging from 3 K to 180 K. Based on the Dulong-Petit law, $C$ is close to the Dulong-Petit limit ($3nR \sim 350$ J mol$^{-1}$ K$^{-1}$, where $R = 8.314$ J mol$^{-1}$ K$^{-1}$, $n$ is the number of atoms per formula, which is 14 for $Ba_5In_4Bi_5$.) at 180 K for the phonon contribution of the sample protected by N-type grease. The specific heat capacity measured without magnetic field shows an anomaly around 4.1 K, as demonstrated in Fig. 4(b), which agrees well with the superconducting transition temperatures in magnetic susceptibility and resistivity. The superconducting transition is totally suppressed under a magnetic field of 50 kOe, which exceeds the estimated upper critical fields of Ba$_5$In$_4$Bi$_5$. The specific heat capacity of the normal state in the temperature range of 3 K to 7 K can be fitted with the Debye model $C/T = \gamma + \beta T^2 + \delta T^4$ (Fig. 4(c)), where the first item is the normal state electronic contribution (Sommerfeld coefficient), the second and third items both derived from the lattice contribution (phonon specific heat coefficient). The fitted parameters are $\gamma$ = 37.4 mJ mol$^{-1}$ K$^{-2}$ and $\beta$ = 11.2 mJ mol$^{-1}$ K$^{-4}$. The Debye temperature $\Theta_D$ can be determined using $[12\pi^4 nR/(5\beta)]^{1/3}$ according to $\beta$, thus being 134 K. We then calculated the electron-phonon coupling constant $\lambda_{ep}$ according to the inverted McMillans formula $\lambda_{ep} = \frac{1.04+\mu^*\ln(\frac{\Theta_D}{1.45T_C})}{(1-0.62\mu^*)\ln(\frac{\Theta_D}{1.45T_C})-1.04}$, where the Coulomb pseudopotential parameter $u^*$ is set to be 0.13 (usually assumed to be 0.1 ~ 0.15 [29,30]). The $\lambda_{ep}$ has been determined to be 0.8, which is smaller than the minimum ($\lambda_{ep}$ = 1) of strong-coupling superconductor. Generally speaking, superconductors with $\lambda_{ep} \geq 1$ are classified as strongly coupled ones, while $0.7 \leq \lambda_{ep} \leq 1$ means moderate coupling [29,31]. This indicates that Ba$_5$In$_4$Bi$_5$ is a moderately coupled superconductor. The electronic specific heat capacity ($C_{el}$) in the superconducting state can be obtained by subtracting the phonon contribution of specific heat capacity in normal state (Fig. 4(d)). By extrapolating the low temperature data of $C_{el}/T$ fitted by the Bardeen-Cooper-Schrieffer (BCS) theory ($C_e/T \propto \exp(-\Delta/k_BT)$, where $k_B$ is the Boltzmann constant, $\Delta$ the SC gap. [32]) down to 0 K, a residual value $\gamma_0$ = 3.82 ± 0.098 mJ mol$^{-1}$ K$^{-2}$ is obtained, indicating a non-superconducting fraction of about 9.9%-10.7%. So the superconducting-state Sommerfeld coefficient ($\gamma_n$) for the crystal is about 33.58 mJ mol$^{-1}$ K$^{-2}$. From Fig. 4(d), the specific heat capacity jump $\Delta C_{el}/T_c$ is about 55.6 mJ mol$^{-1}$ K$^{-2}$. The derived $\Delta C_{el}/\gamma_n T_c = 1.65$, larger than that (1.426) of weak-coupling BCS-type superconductor, which agrees with the moderate coupling shown by $\lambda_{ep}$. The apparent superconducting fraction (89.3%-90.1%) determined by specific heat capacity and the non-saturated ZFC data at 3 K (Fig. 2(a)), provide the convinced evidence for bulk SC in Ba$_5$In$_4$Bi$_5$. Combining the results presented above, the noninteracting DOS at the Fermi level $N(E_F) = 8.5$ states eV$^{-1}$ per formula unit is obtained from $N(E_F) = 3\gamma/[\pi^2 k_B^2(1+\lambda_{ep})]$, which agrees with the results obtained from the calculation. The larger DOS at the Fermi level may be one of the reasons for the SC of Ba$_5$In$_4$Bi$_5$.

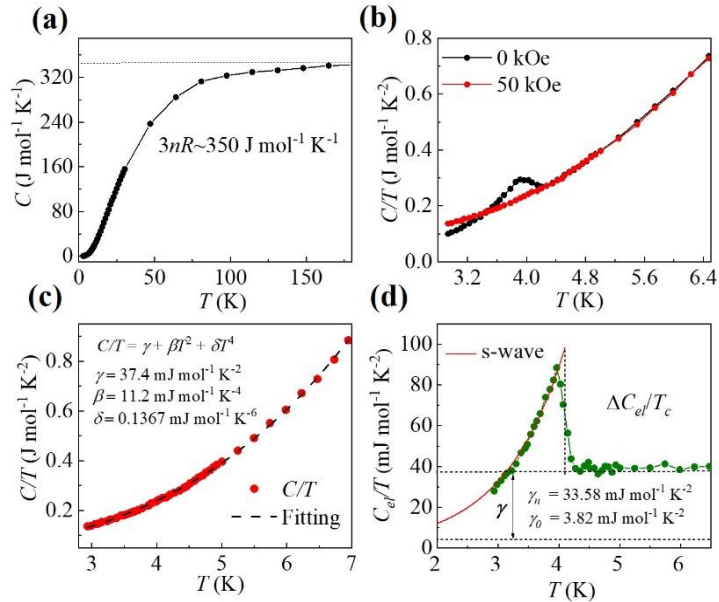

**FIG. 4.** (a) The specific heat capacity of Ba$_5$In$_4$Bi$_5$ single crystal as a function of temperature under zero magnetic field. (b) The $C/T$ as a function of temperature in the low temperature range under the magnetic field of 0 kOe and 50 kOe for $H \perp (001)$ plane. (c) The $C/T$ as a function of temperature at low temperature under a magnetic field of 50 kOe for $H \perp (001)$ plane. The black dashed line is the fitting by using the

Debye model $C/T = \gamma + \beta T^2 + \delta T^4$. (d) The electronic term in 0 kOe obtained after subtracting the phonon term determined in 50 kOe. The red solid line shows the calculated $C_{el}/T$ curve by the BCS model.

The careful magnetism, resistivity, and specific heat capacity measurements and first-principles calculations clearly show the coexisting inherent SC and nontrivial topological band structure in $Ba_5In_4Bi_5$. The calculations indicate that the nodal chain in $Ba_5In_4Bi_5$ is close to the Fermi energy, which implies that the nodal chain should be detectable for different experimental techniques on band structure features. Further characterizations, like angle-resolved photoemission spectroscopy and scanning tunneling microscopy/spectroscopy, are needed to understand the interplay of SC and nontrivial band topology in $Ba_5In_4Bi_5$.

In conclusion, our results demonstrate that $Ba_5In_4Bi_5$ is a noncentrosymmetric topological nodal chain semimetal with intrinsic SC. The band structure calculations show the nodal loops around the high-symmetry R and X points form a nodal chain by vertically connecting each other. We first observe the bulk SC with $T_c$ of 4.1 K in $Ba_5In_4Bi_5$, which is determined to be a type-II superconductor. The value of $\lambda_{ep}$ (0.8) and $\Delta C_{el}/\gamma_n T_c$ (1.65) suggest $Ba_5In_4Bi_5$ should be ascribed to a moderately coupled superconductor. The results show that $Ba_5In_4Bi_5$ is a good platform to investigate the interplay of SC and topology.

**Acknowledgements**

Y. Z. Ma, Y. L. Wang, and G. Wang would like to thank Profs. X. L. Chen, J. P. Hu, X. J. Zhou, and G. D. Liu of the Institute of Physics, Chinese Academy of Sciences for helpful discussions. This work was partially supported by the National Key Research and Development Program of China (2018YFE0202602 and 2017YFA0302902) and the National Natural Science Foundation of China (51832010). Work at Ames National Laboratory was supported by the U.S. Department of Energy, Office of Basic Energy Science, Division of Materials Sciences and Engineering. Ames Laboratory is operated for the U.S. Department of Energy by Iowa State University under Contract No. DE-AC02-07CH11358.

## Supplemental Material

### Experiment and Methods

**Single crystal growth**

Ba$_5$In$_4$Bi$_5$ single crystals were grown by a flux method with excess In as flux. Starting materials of as-received Ba lump (99.9%), In shot (99.999%), and Bi shot (99.999%), all from Alfa Aesar, were weighed with a molar ratio of 1.02:3:1, put into a fritted alumina crucible set (Canfield Crucible Set or CCS) [1]. All manipulations were carried out in an argon-filled glovebox. The alumina crucible set was then sealed in a fused silica tube under partial argon pressure (0.1 atm). The ampule was heated to 1173 K with a ramp rate of 200 K/h, dwelled for 6 h, then cooled to 1073 K in 2 h and kept for 6 h, then slowly cooled down to 823 K over a period of 5 d, at which temperature the excess liquid was removed by a centrifuging process [2]. The as-grown Ba$_5$In$_4$Bi$_5$ single crystals with silver gray color show typical size up to 2.5 × 1.6 × 0.8 mm$^3$, as displayed in Fig. S1(a).

**Single crystal and powder X-ray diffraction (PXRD)**

A single crystal was selected and cut in the glovebox and protected with N-type grease. The single crystal X-ray diffraction data of Ba$_5$In$_4$Bi$_5$ were collected by Bruker D8 VENTURE single crystal diffractometer with multilayer monochromatized Mo Kα radiation ($\lambda$ = 0.71073 Å) at room temperature (293 K), and the crystal structure was solved by direct methods and refined by the full-matrix least-squares method on F$^2$ with SHELXL software [3,4]. The as-grown single crystals were ground in the argon-filled glove box and protected with the sealed sample platform of X-ray diffraction instrument.

The PXRD measurement was carried out using a PANalytical X'Pert PRO diffractometer (Cu Kα radiation, $\lambda$ = 1.54178 Å) operating at 40 kV voltage and 40 mA current at 293 K [5].

**Chemical composition determination**

Elemental analyses were carried out by using (1) a scanning electron microscope (SEM, Hitachi S-4800) equipped with energy-dispersive X-ray spectroscopy (EDX). For each of the two crystals that came from one batch, 5 spots in different areas were measured, and the exposed time of samples in air is less than 5 minutes. The composition homogeneity of the sample was checked by the elemental mapping; (2) inductively coupled plasma-atomic emission spectroscopy (ICP-AES) (Teledyne Leeman Laboratories Prodigy 7). For each batch, the ICP-AES measurement was performed on four single crystals.

**Crystal orientation**

Crystal orientation of $Ba_5In_4Bi_5$ crystal was determined by using the single crystal Laue diffractometer (Huber Diffraktionstechnik GmbH & Co. KG V42833-3). The largest flat surface is identified as the (001) plane based on the successful indexing of corresponding diffraction images, meaning that the largest surface is perpendicular to the $c$ axis.

**Physical property measurements**

Resistivity and specific heat capacity measurements were carried out on Physical Property Measurement System (PPMS, Quantum Design), and the magnetic susceptibility was measured using PPMS equipped with Vibrating Sample

Magnetometer. Magnetic susceptibility under a fixed magnetic field of 10 Oe was obtained in both zero-field cooling (ZFC) and field cooling (FC) modes running demagnetization sequence. The fluctuation of the magnetic field was ±3 Oe originated from the difference between the actual value and the value indicated by the instrument. Isothermal magnetization curve was measured by sweeping the magnetic field under various temperatures. Resistivity was measured by attaching platinum wires to a crystal with silver paste using a standard four-probe dc technique. The samples were handled in the argon-filled glove box. Moreover, the electrodes were covered by a layer of low-temperature glue (GE 7031 Varnish) to prevent sample degradation. The contact resistance is smaller than 5 Ω. The specific heat capacity was measured from 3 K-180 K in zero field (for the superconducting state) and under 50 kOe magnetic field (for the normal state) applied along the $c$ axis by using the thermal relaxation-time method.

**First-principles calculations**

The electronic band structure of $Ba_5In_4Bi_5$ was calculated in the framework of density functional theory, as implemented using the projector augmented wave method encoded in the Vienna Ab initio Simulation Package [6-8]. The generalized gradient approximation of the Perdew-Burke-Ernzerhof type was adopted for exchange-correlation functional [9]. The plane-wave cutoff energy was set to be 500 eV and with a 5 × 5 × 6 Γ-centered $k$ points in self-consistent calculation. The energy convergence accuracy was chosen to be $10^{-8}$ eV and the spin-orbit coupling (SOC) was considered in this calculation. In addition, the tetrahedron method was employed in the calculation of density of states (DOS) to ensure the smoothness of DOS. In nodal line calculation, a denser 401 × 401 $k$ points mesh grid in plane was adopt to acquire a high accuracy. If the energy difference between the two energy bands at a certain $k$ point was less than 0.003 eV, it was considered to be a band intersection.

**Table S1**. Crystallographic data and structure refinement parameters for $Ba_5In_4Bi_5$.

| Formula | $Ba_5In_4Bi_5$ (this work) | $Ba_5In_4Bi_5$ (Ref.) [10] |
|---|---|---|
| Formula weight | 2190.88 | 2190.88 |
| $T$ (K) | 293(2) | 293(2) |
| Space group | *P4nc* | *P4nc* |
| Unit cell dimensions | $a$ = 10.6132(6) Å | $a$ = 10.620(1) Å |
|  | $c$ = 9.0174(11) Å | $c$ = 9.009(2) Å |
| Volume | 1015.72(17) Å$^3$ | 1016.0(3) Å$^3$ |
| Z | 2 | 2 |
| Density (calculated) | 7.163 g/cm$^3$ | 7.161 g/cm$^3$ |
| Goodness-of-fit on F$^2$ | 1.085 | 1.219 |
| Final R indices [I > 2sigma(I)] | R1 = 0.0341, wR2 = 0.0812 | R1 = 0.0237, wR2 = 0.0700 |
| R indices (all data) | R1 = 0.0348, wR2 = 0.0817 | R1 = 0.0273, wR2 = 0.0843 |

**Table S2.** Atomic coordinates and equivalent isotropic displacement parameters for $Ba_5In_4Bi_5$. $U_{eq}$ is defined as one third of the trace of the orthogonalized $U_{ij}$ tensor.

| atom | Wyckoff | x | y | z | $U_{eq}$ (Å$^2$) |
|---|---|---|---|---|---|
| Bi1 | 8c | 0.1094(1) | 0.3092(1) | 0.0103(3) | 0.013(1) |
| Bi2 | 2a | 0 | 0 | 0.3822(3) | 0.008(1) |
| Ba1 | 8c | 0.3115(1) | 0.1060(1) | 0.3263(3) | 0.012(1) |
| Ba2 | 2a | 0 | 0 | 0 | 0.006(1) |
| In1 | 8c | 0.3168(1) | 0.4281(1) | 0.1686(3) | 0.016(1) |

**Table S3.** Selected interatomic distances [Å] in $Ba_5In_4Bi_5$.

| | | | |
|---|---|---|---|
| Bi2-Ba1 | 3.5284 (10) × 4 | Bi1-In1 | 2.9106 (14) |
| Bi2-Ba2 | 3.4470 (3) | Ba1-In1 | 4.0823 (19) |
| Bi2-In1 | 3.3212 (19) × 4 | Ba1-In1 | 3.8660 (19) |
| Bi1-Ba1 | 3.6526 (14) | Ba1-In1 | 3.7032 (16) |
| Bi1-Ba1 | 3.6828 (12) | Ba1-In1 | 3.8301 (16) |
| Bi1-Ba1 | 4.1678 (13) | Ba2-Bi1 | 3.4825 (6) |
| Bi1-Ba1 | 3.6772 (12) | Ba2-In1 | 3.6460 (2) × 4 |
| Bi1-Ba1 | 3.6574 (12) | In1-In1 | 2.9540 (17) × 2 |
| Bi1-In1 | 3.3830 (13)[1] | | |

[1] Intercluster bonds.

**Table S4.** Anisotropic displacement parameters [Å$^2$] for $Ba_5In_4Bi_5$. The anisotropic displacement factor exponent takes the form: $-2p^2 [h^2 a^{*2} U^{11} + ... + 2 h k a^* b^* U^{12}]$.

| | $U^{11}$ | $U^{22}$ | $U^{33}$ | $U^{23}$ | $U^{13}$ | $U^{12}$ |
|---|---|---|---|---|---|---|
| Bi1 | 0.010(1) | 0.014(1) | 0.015(1) | -0.001(1) | 0(1) | -0.004(1) |
| Bi2 | 0.009(1) | 0.009(1) | 0.005(1) | 0 | 0 | 0 |
| Ba1 | 0.013(1) | 0.011(1) | 0.014(1) | 0.001(1) | 0(1) | -0.003(1) |
| Ba2 | 0.007(1) | 0.007(1) | 0.005(1) | 0 | 0 | 0 |
| In | 0.016(1) | 0.012(1) | 0.019(1) | -0.002(1) | -0.010(1) | 0(1) |

**Table S5**. The EDX analyses for Ba$_5$In$_4$Bi$_5$ single crystals.

| No. | Point | Ba atomic ratio (%) | In atomic ratio (%) | Bi atomic ratio (%) |
|---|---|---|---|---|
| 1 | ① | 34.72 | 28.72 | 36.56 |
|   | ② | 36.45 | 27.96 | 35.59 |
|   | ③ | 35.64 | 28.34 | 36.02 |
|   | ④ | 35.98 | 28.45 | 35.57 |
|   | ⑤ | 36.21 | 29.18 | 34.61 |
| 2 | ① | 35.92 | 28.76 | 35.32 |
|   | ② | 34.99 | 28.98 | 36.03 |
|   | ③ | 36.15 | 29.12 | 34.73 |
|   | ④ | 36.29 | 27.98 | 35.73 |
|   | ⑤ | 37.01 | 27.57 | 35.42 |

**Table S6**. Atomic ratio of Ba, In, and Bi in Ba$_5$In$_4$Bi$_5$ single crystals determined by ICP-AES.

| No. | Ba | In | Bi |
|---|---|---|---|
| 1 | 35.52 | 28.90 | 35.58 |
| 2 | 35.64 | 28.76 | 35.6 |
| 3 | 35.81 | 28.49 | 35.7 |
| 4 | 35.43 | 28.83 | 35.74 |

**Table S7**. The upper critical field $H_{c2}(0)$ estimated from empirical equation, Werthamer-Helfand-Hohenberg (WHH) model, and Ginzburg-Landau (GL) model.

| $H_{c2}(0)$ | Units | A | $H \,/\!/\, (001)$ | $H \perp (001)$ |
|---|---|---|---|---|
| empirical equation | kOe |  | 36.3 | 25.5 |
| WHH model | kOe | -0.69 | 24.1 | 16.8 |
|  |  | -0.73 | 25.5 | 17.8 |
| GL model | kOe |  | 34.5 | 23.5 |

Table S8. The superconducting parameters of $Ba_5In_4Bi_5$ single crystal.

| Parameters | Units | $H \parallel (001)$ | $H \perp (001)$ |
|---|---|---|---|
| $T_c$ | K | 4.1 | |
| RRR | | 6 | |
| $H_{c1}(0)$ | Oe | 92 | 82 |
| $H_{c2}(0)$ | kOe | 36.3 | 25.5 |
| $\xi(0)$ | nm | 11 | 8 |
| $\kappa(0)$ | | 25 | 22 |
| $\lambda(0)$ | nm | 242 | 227 |
| $\Delta C/\gamma T_c$ | | 1.65 | |
| $\Theta_D$ | K | 134 | |
| $\lambda_{ep}$ | | 0.8 | |

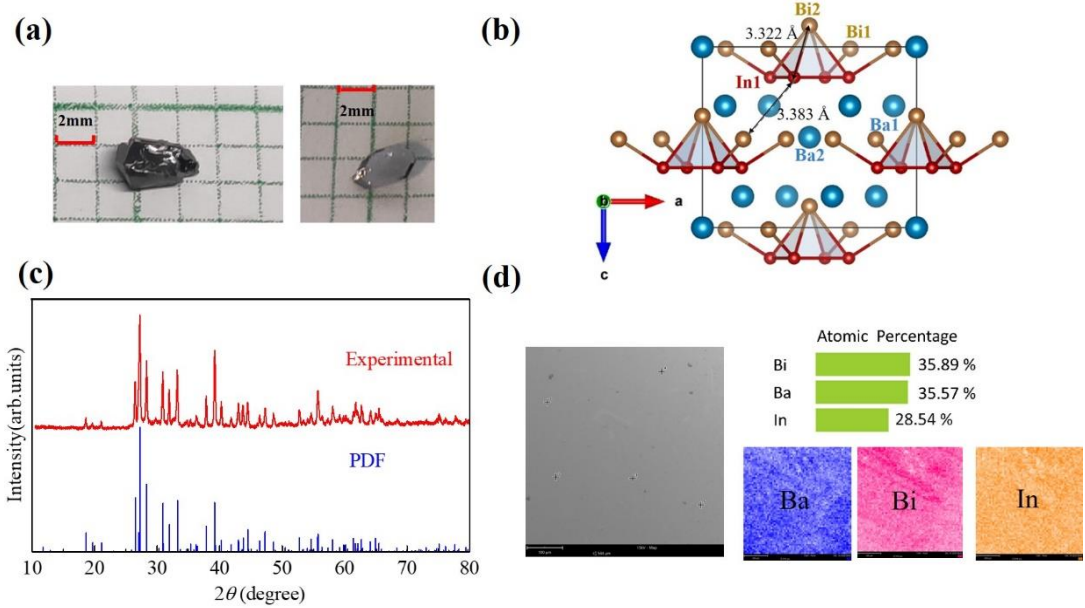

**Fig. S1.** (a) The optical images of $Ba_5In_4Bi_5$ single crystals. The grid size is 2 mm. (b) Crystal structure of $Ba_5In_4Bi_5$ viewed along the *b* axis, and the square pyramidal clusters are connected by Bi1-In1 from different pyramids. (c) The PXRD pattern of crushed $Ba_5In_4Bi_5$ single crystals. (d) The EDX spectrum and elemental mapping of $Ba_5In_4Bi_5$ single crystal.

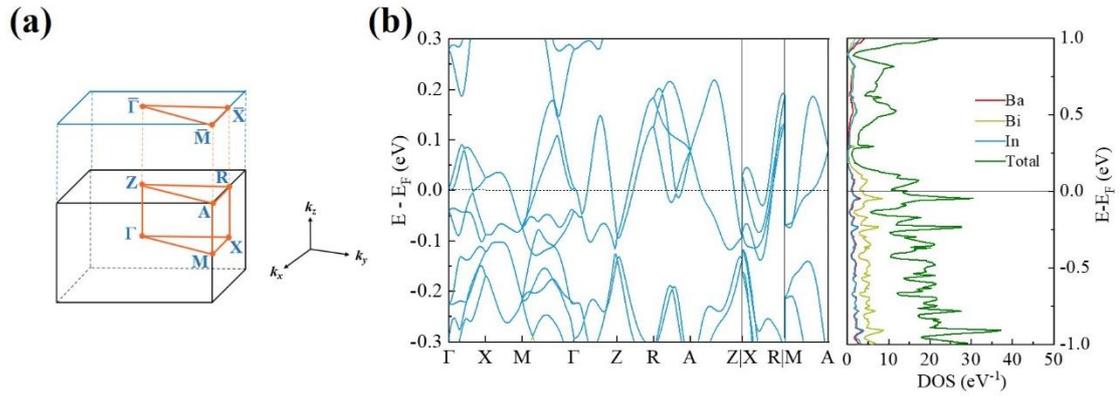

**Fig. S2.** (a) The bulk and (001)-projected surface Brillouin zones of $Ba_5In_4Bi_5$. (b) The calculated electronic band structure and DOS of $Ba_5In_4Bi_5$ with SOC.

Figure S2(b) presents the metallic character of $Ba_5In_4Bi_5$ with several electron-like bands crossing the Fermi level, and the contribution of Bi is dominant near the Fermi level in the energy range from -1 eV to 0 eV, as shown by the DOS of $Ba_5In_4Bi_5$ on the right panel of Fig. S2(b).

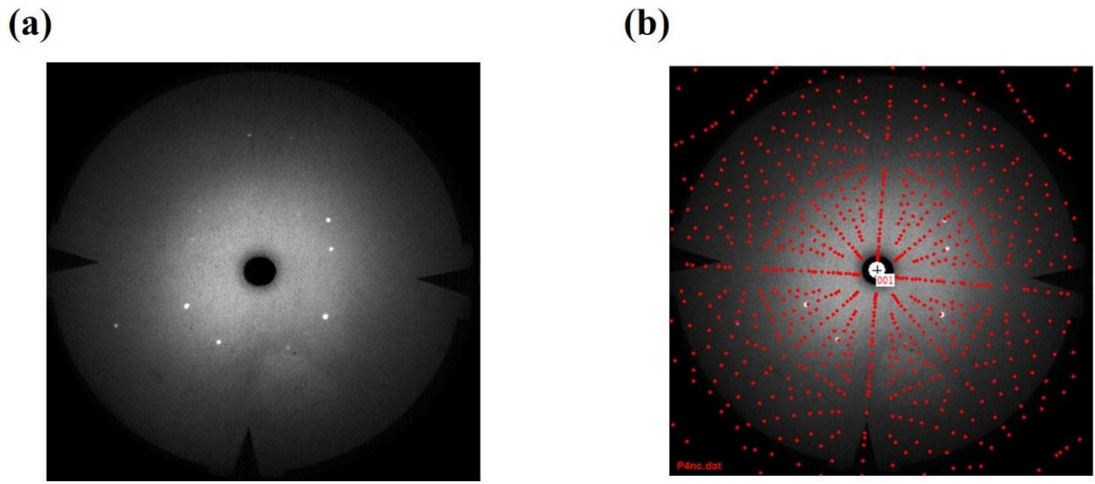

**Fig. S3.** (a) (b) The Laue diffraction pattern of (001) surface for $Ba_5In_4Bi_5$ single crystal. The scarcity of points may be due to the surface oxidation of the sample.

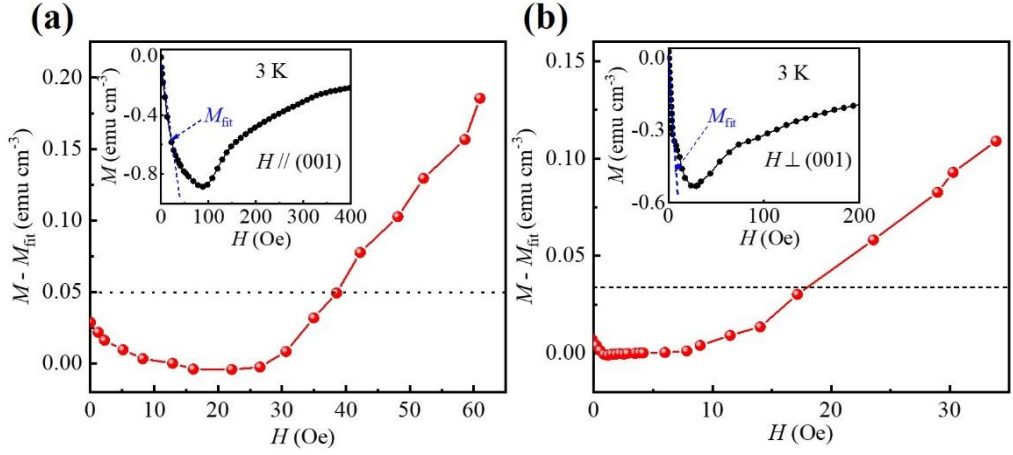

**Fig. S4.** (a) (b) The field dependence of the difference between magnetization (*M*) and the low-field linear fit to the magnetization ($M_{\text{fit}}$) at 3 K for *H* // (001) and *H* ⊥ (001). The dashed lines are guides used for obtaining $H_{c1}^*$. The insets show the field-dependent magnetization curves at 3 K for *H* // (001) and *H* ⊥ (001).

Figure S4(a) and (b) presents the difference between *M* measured at 3 K and the $M_{\text{fit}}$ (shown as the blue dash lines in the inset of Fig. S4(a) and (b)) in the low magnetic field range for *H* // (001) and *H* ⊥ (001). The lower critical field ($H_{c1}^*$) is the field where the deviation from linear behavior ($\Delta M = M - M_{\text{fit}}$) is equal to 0.05 emu cm$^{-3}$ for *H* // (001) or 0.03 emu cm$^{-3}$ for *H* ⊥ (001), respectively. (Note: $\Delta M$ is about 5% of the magnetization value *M* for applied field of $H_{c1}^*$ obtained at *T* = 3 K for *H* // (001) and *H* ⊥ (001).)

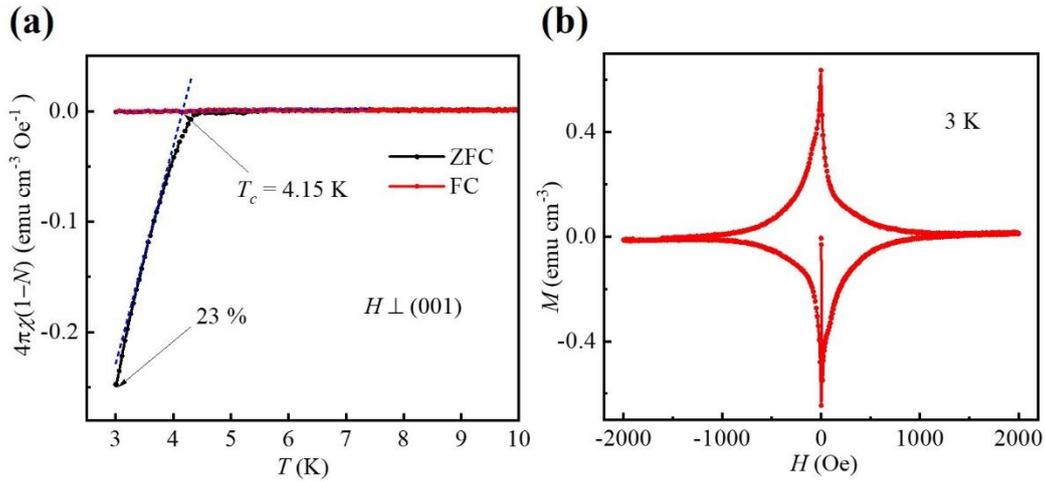

**Fig. S5.** (a) The ZFC and FC magnetic susceptibility for $Ba_5In_4Bi_5$ single crystal under the magnetic field of 10 Oe for $H \perp (001)$ plane after magnetic susceptibility measurement for $H // (001)$. (b) The magnetic hysteresis loop of the same crystal measured at 3 K.

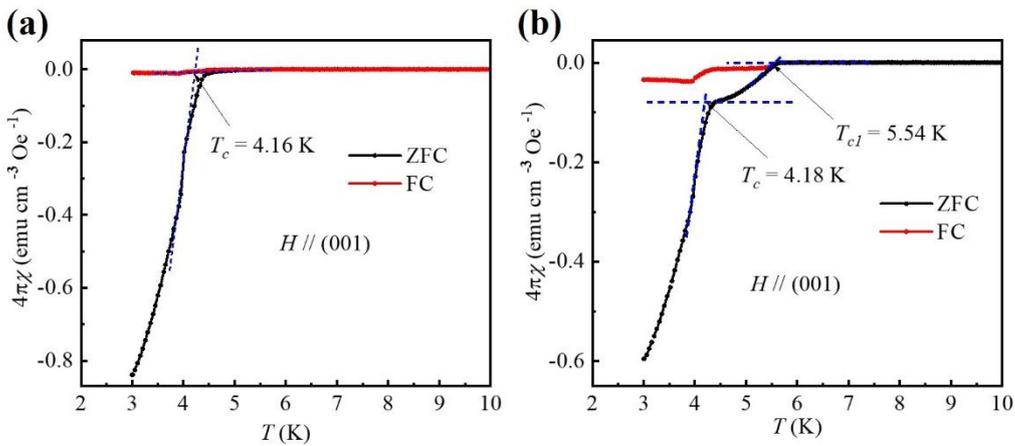

**Fig. S6.** (a) (b) The magnetic susceptibility of an as-grown $Ba_5In_4Bi_5$ single crystal and the same crystal exposed in air for two days (The demagnetization effect is not considered).

Figure S6 displays the temperature dependence of magnetic susceptibility of an as-grown $Ba_5In_4Bi_5$ single crystal and the same crystal exposed in air for two days. The as-grown $Ba_5In_4Bi_5$ single crystal shows a $T_c$ about 4.16 K with superconducting shielding volume fraction near 83% at 3 K. After exposed in air for two days, another superconducting transition at 5.54 K emerges, mostly due to the partial decomposition of $Ba_5In_4Bi_5$ in air.

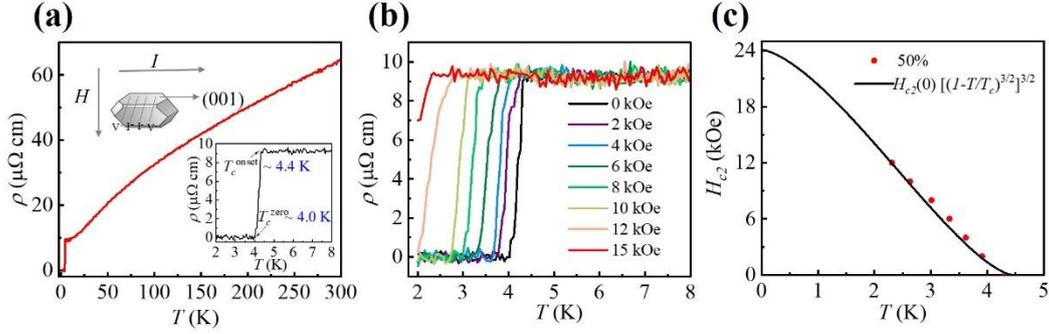

**Fig. S7.** (a) The temperature dependence of resistivity for a $Ba_5In_4Bi_5$ single crystal with current applied in (001) plane for $H \perp$ (001) plane. The inset displays the enlarged resistivity curve at low temperature range. (b) The resistivity as a function of temperature under different magnetic fields for $H \perp$ (001) plane. (c) The temperature dependence of the upper critical field. The solid line shows fitting by the equation $H_{c2}(T) = H_{c2}(0)[1-(T/T_c)^{3/2}]^{3/2}$.

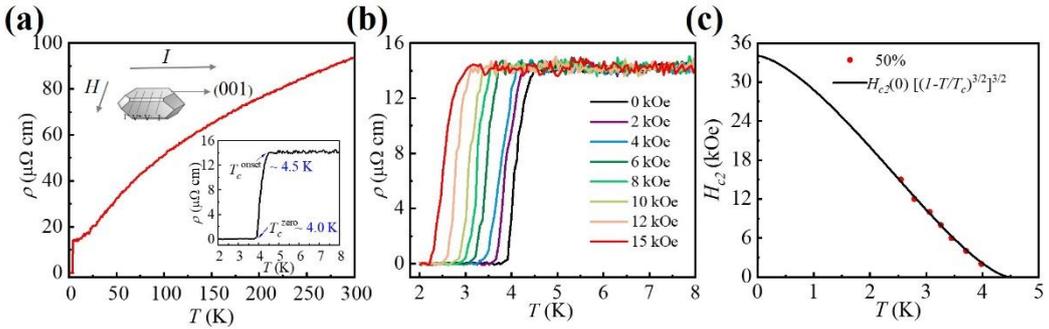

**Fig. S8.** (a) The temperature dependence of resistivity for a $Ba_5In_4Bi_5$ single crystal with current applied in (001) plane for $H //$ (001) plane. The inset displays the enlarged resistivity curve at low temperature range. (b) The resistivity as a function of temperature under different magnetic fields for $H //$ (001) plane. (c) The temperature

dependence of the upper critical field. The solid line shows fitting by the equation $H_{c2}(T) = H_{c2}(0)[1-(T/T_c)^{3/2}]^{3/2}$.

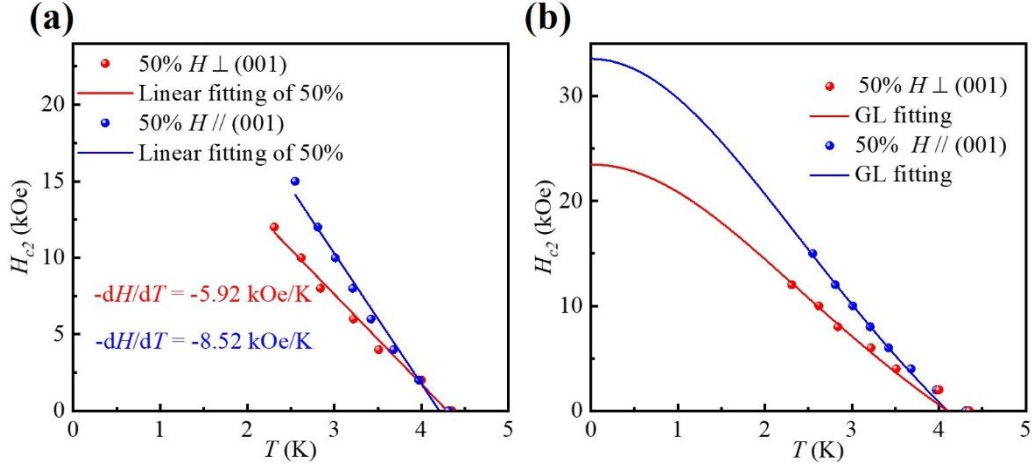

**Fig. S9.** (a), (b) The temperature dependence of the upper critical field for $H \perp (001)$ plane and $H // (001)$ plane, respectively. The solid lines show (a) the linear fittings with $dH_{c2}/dT|_{(T=T_c)}$ and (b) the fittings by GL model, respectively.

Figure S9(a) shows a linear relationship between $H_{c2}(T)$ and $T_c$ for $H \perp (001)$ and $H // (001)$ plane, respectively, which can be used to calculate $H_{c2}(0)$ according to the WHH expression $H_{c2}(0) = -AT_c dH_{c2}/dT|_{(T=T_c)}$, where A is either 0.69 or 0.73 for the dirty or clean limits, respectively [11,12]. Except for this, we fit the data shown in Fig. S9(b) with GL equation $H_{c2}(T) = H_{c2}(0)(1 - t^2)/(1 + t^2)$ [13], where $t = T/T_c$, which yields a value that agrees with the calculated value using the empirical equation, as shown in Table S7. The difference of $H_{c2}(0)$ obtained by the WHH equation is mainly due to the upper critical field is not completely linear with temperature.

We employ $H_{c2}(0) = 25.5$ kOe ($H \perp (001)$) and 36.3 kOe ($H // (001)$) for further calculations. According to the GL theory, the superconducting coherent length ($\xi$) can be obtained by the equations (1) and (2):

$$H_{c2}^{c}(0) = \Phi_0/(2\pi\xi_{ab}^2(0)) \tag{1}$$

$$H_{c2}^{ab}(0) = \Phi_0/(2\pi\xi_{ab}(0)\xi_c(0)) \tag{2}$$

where $\Phi_0$ is the quantum of flux, $H_{c2}^{c}$ and $H_{c2}^{ab}$ the corresponding upper critical fields for $H \perp$ (001) plane and $H //$ (001) plane, respectively. The GL parameters $\kappa_{ab}(0)$ and $\kappa_c(0)$ are given by equations (3) and (4):

$$H_{c2}^{ab}(0)/H_{c1}^{ab}(0) = 2\kappa_{ab}^2(0)/\ln\kappa_{ab}(0) \tag{3}$$

$$H_{c2}^{c}(0)/H_{c1}^{c}(0) = 2\kappa_{c}^2(0)/\ln\kappa_{c}(0) \tag{4}$$

In addition, based on the relation between the coherence length and GL parameter,

$$\kappa_{ab}(0) = \lambda_{ab}(0)/\xi_c(0) = [\lambda_{ab}(0)\lambda_c(0)/\xi_{ab}(0)\xi_c(0)]^{1/2}$$

the magnetic penetration depth can be obtained by the equation (5):

$$\kappa_c(0) = \lambda_{ab}(0)/\xi_{ab}(0) \tag{5}$$

where $\xi_{ab}/\xi_c = \lambda_c/\lambda_{ab}$, all the calculated parameters are listed in Table S8.